\documentclass{PoS}

\title{The QCD equation of state and the effects of the charm}

\ShortTitle{The QCD equation of state and the effects of the charm}

\author{Szabolcs~Borsanyi$^a$, Gergely Endrodi$^b$, Zoltan~Fodor$^{a,c,d}$, Sandor~D.~Katz$^{a,d}$, \speaker{Stefan~Krieg}$^{a,c}$, Claudia Ratti$^e$, Chris Schroeder$^{a,f}$, Kalman~K.~Szabo$^a$\\
\llap{$^a$}Bergische Universit\"at Wuppertal, D-42119 Wuppertal, Germany\\
\llap{$^b$}Universit\"at Regensburg, D-93040 Regensburg, Germany\\
\llap{$^c$}IAS, J\"ulich Supercomputing Centre, Forschungszentrum J\"ulich, D-52425 J\"ulich, Germany\\
\llap{$^d$}Institute for Theoretical Physics, E\"otv\"os University, H-1117 Budapest, Hungary\\
\llap{$^e$}Universit\`a degli Studi di Torino and INFN, Sezione di Torino, I-10125 Torino, Italy\\
\llap{$^f$}Lawrence Livermore National Laboratory, Livermore, California 94550, USA\\
E-mail: \email{s.krieg@fz-juelich.de}
}

\abstract{We present an update on the QCD equation of state of the Wuppertal-Budapest Collaboration, extending our previous studies \cite{Aoki:2005vt, Borsanyi:2010cj}. A Symanzik improved gauge and a stout-link improved staggered fermion action is utilized. We discuss partial quenching and present preliminary results for the fully dynamical charmed equation of state.
}

\FullConference{XXIX International Symposium on Lattice Field Theory \\
                 July 10 - 16 2011\\
                 Squaw Valley, Lake Tahoe, California}

\begin{document}

\section{Introduction}
The properties of the Quark-Gluon-Plasma (QGP) are the focus of a large number of heavy ion experiments, such as ALICE at LHC, CERN SPS, RHIC at BNL, and, in the future, FAIR at GSI Darmstadt. Hydrodynamical models, based on the observation that the QGP is well approximated by an ideal liquid
\cite{Teaney:2000cw, Teaney:2001av, Kolb:2003dz}, 
require as an input parameter the equation of state (EOS) of Quantum Chromodynamics (QCD). 

Computing the EOS from QCD directly is possible through simulations of the lattice regularized theory, Lattice QCD. As algorithms and computers improve, such simulations aiming to understand the features of the QGP are reaching unprecedented levels of accuracy both for mapping the phase diagram~\cite{Fodor:2001au, Fodor:2004nz, Aoki:2006we, Endrodi:2011gv, Bali:2011qj} or studying bulk obervables~\cite{ Aoki:2005vt, Borsanyi:2010cj, Aoki:2006br, Aoki:2009sc, Borsanyi:2010bp,  Bazavov:2009mi, Bazavov:2010pg, Bazavov:2011nk,  Levkova:2012jd}. 
Here, we provide a status report on our efforts to improve upon our result of \cite{Borsanyi:2010cj} for the $N_f=2+1$ EOS 
by performing a controlled continuum limit extrapolation and by studying the effects of an additional sea charm quark ($N_f=2+1+1$ EOS). 

\section{Continuum estimate for the $N_f=2+1$ EOS}
\begin{figure}
\begin{center}
\includegraphics*[width=.65\textwidth]{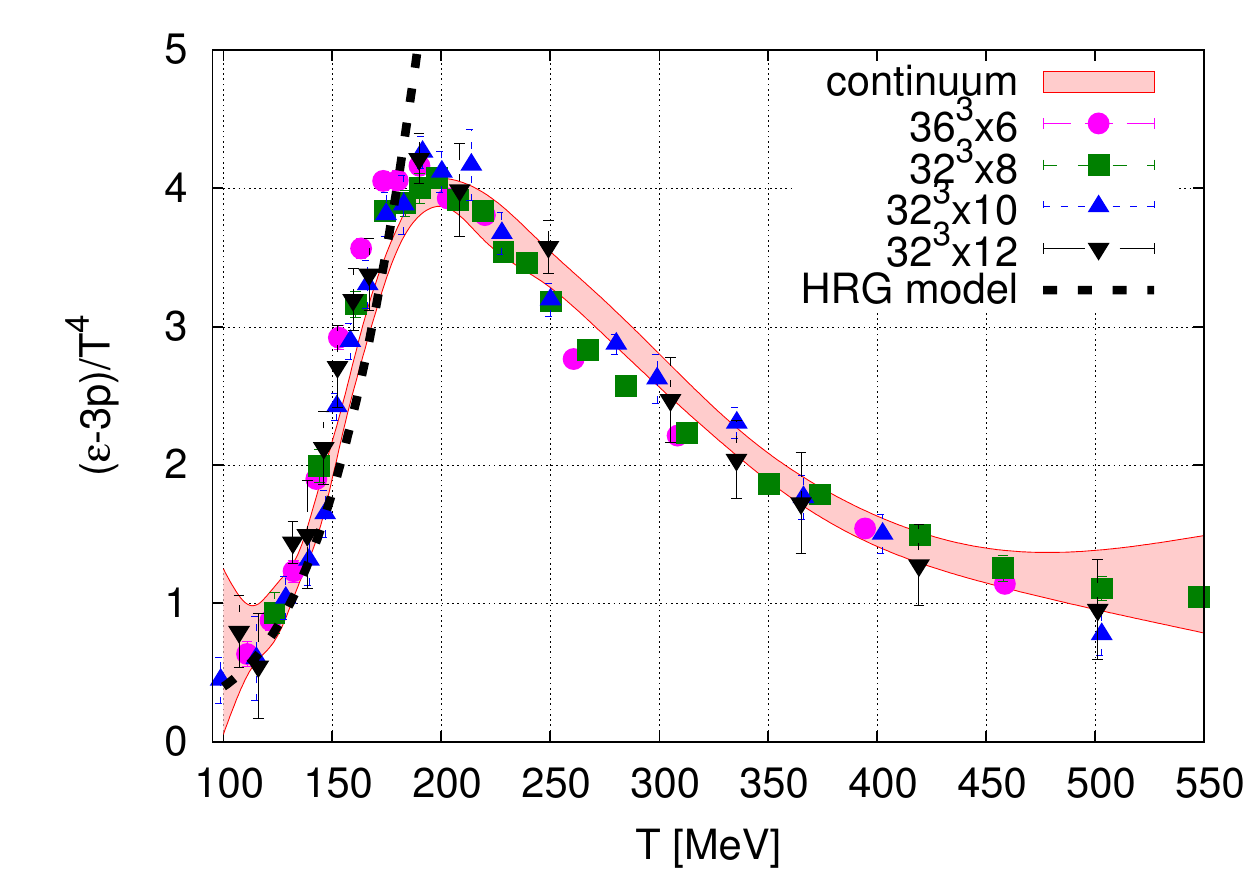}
\end{center}
\caption{Continuum limit estimate for the $N_f=2+1$ QCD equation of state. No tree level improvement was applied to the data points prior to computing the continuum estimate.  
Perfect agreement is found with the Hadron Resonance (HRG) model using physical parameters.
Further data is needed for a complete continuum limit result.}
\label{fig:cont}
\end{figure}
In order to perform a continuum extrapolation for the EOS, we interpolated the results for any given $N_t$ using a standard cubic spline ansatz. Furthermore, we augmented our dataset from \cite{Borsanyi:2010cj} by adding a set of $N_t=12$ points, thereby making a spline interpolation possible. We then varied the details of the spline interpolation to account for the statistical uncertainties and, by adding or removing spline points, to estimate the systematic effects due to this particular interpolation procedure. The continuum limit was then taken with an $O(a^2)$ fit ansatz. The result for the trace anomaly is depicted in Figure~\ref{fig:cont}. While the data points depicted in Figure~\ref{fig:cont} have been corrected for tree level effects, no such correction was applied prior to computing the continuum extrapolation. As already discussed in \cite{Borsanyi:2010cj}, these improvement factors would not influence the continuum extrapolation procedure. This is also nicely visible in Figure~\ref{fig:cont}, since the tree level improved data points agree perfectly with the continuum estimate. Furthermore, excellent agreement of the lattice data with the prediction of the Hadron Resonance Gas (HRG) model was found, strengthening our assessment that the results of \cite{Borsanyi:2010cj} are accurate.

\begin{figure}
\begin{center}
\includegraphics*[width=.65\textwidth]{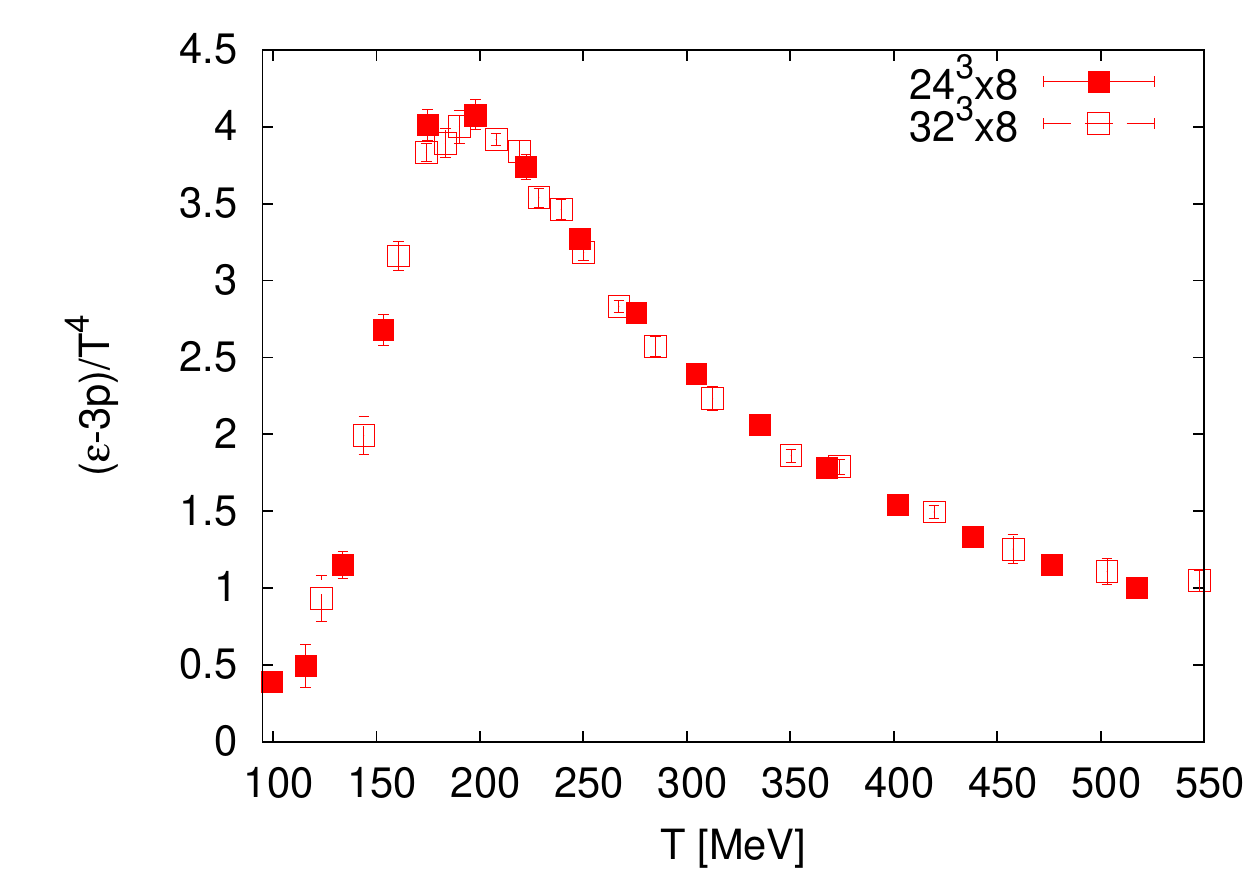}
\end{center}
\caption{Finite volume study for our EOS. Shown is the trace anomaly for two different volumes at $N_t=8$. Complete agreement between the two volumes is found, extending our evidence in \cite{Borsanyi:2010cj}, where the same outcome was found for $N_t=6$. }
\label{fig:vol}
\end{figure}
Finally, in order to check for finite size effects, we have repeated our analysis at $N_t=8$ with two different aspect ratios, shown in Figure~\ref{fig:vol}. As in \cite{Borsanyi:2010cj}, where the same analysis was done for $N_t=6$, the results agree perfectly. Therefore, we believe our volumes are more than adequate and our results are free of finite size effects.

\section{Partially quenched and dynamical charm effects on the equation of state}
\begin{figure}
\begin{center}
\includegraphics*[width=.40\textwidth]{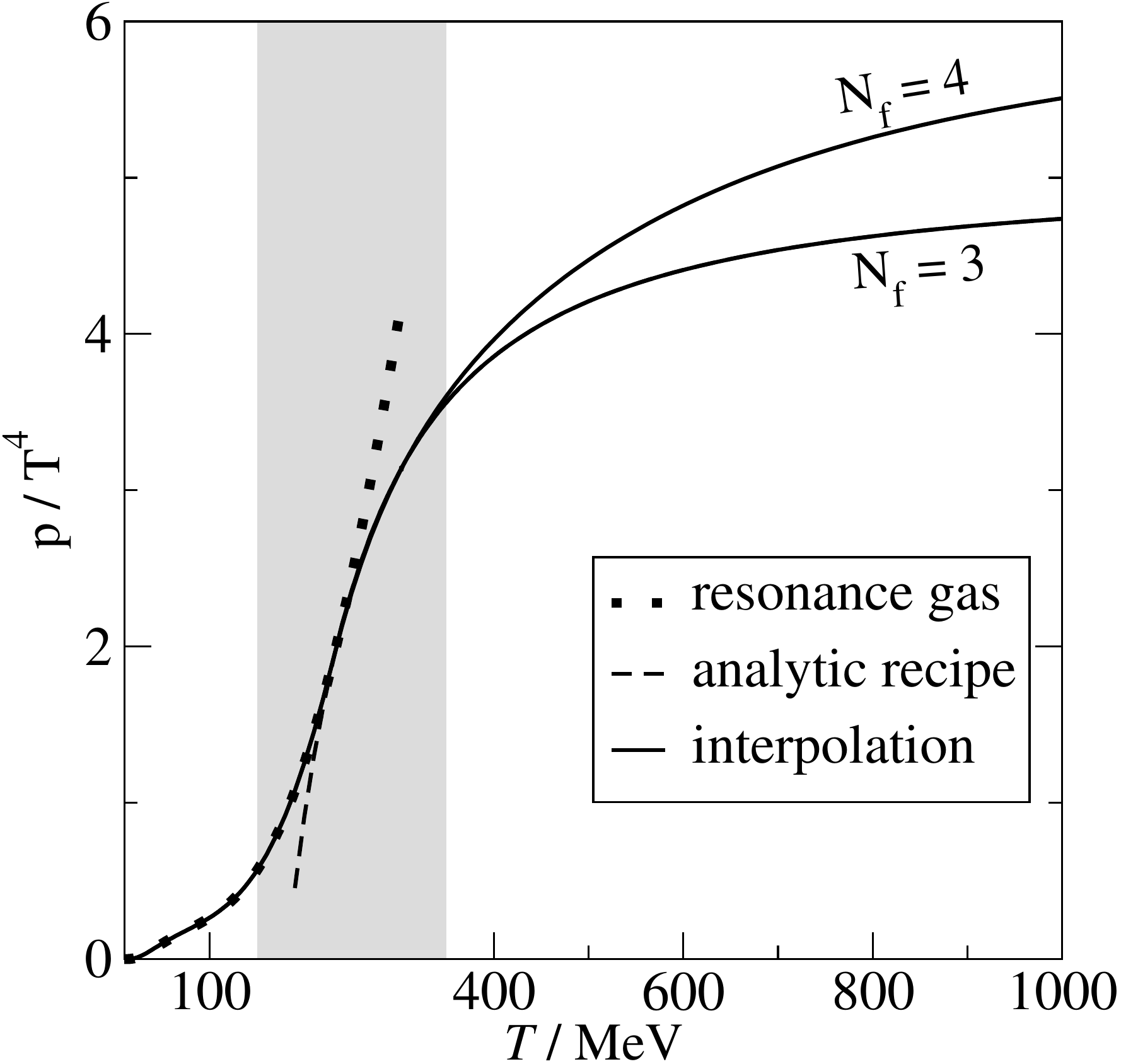}
\includegraphics*[width=.55\textwidth]{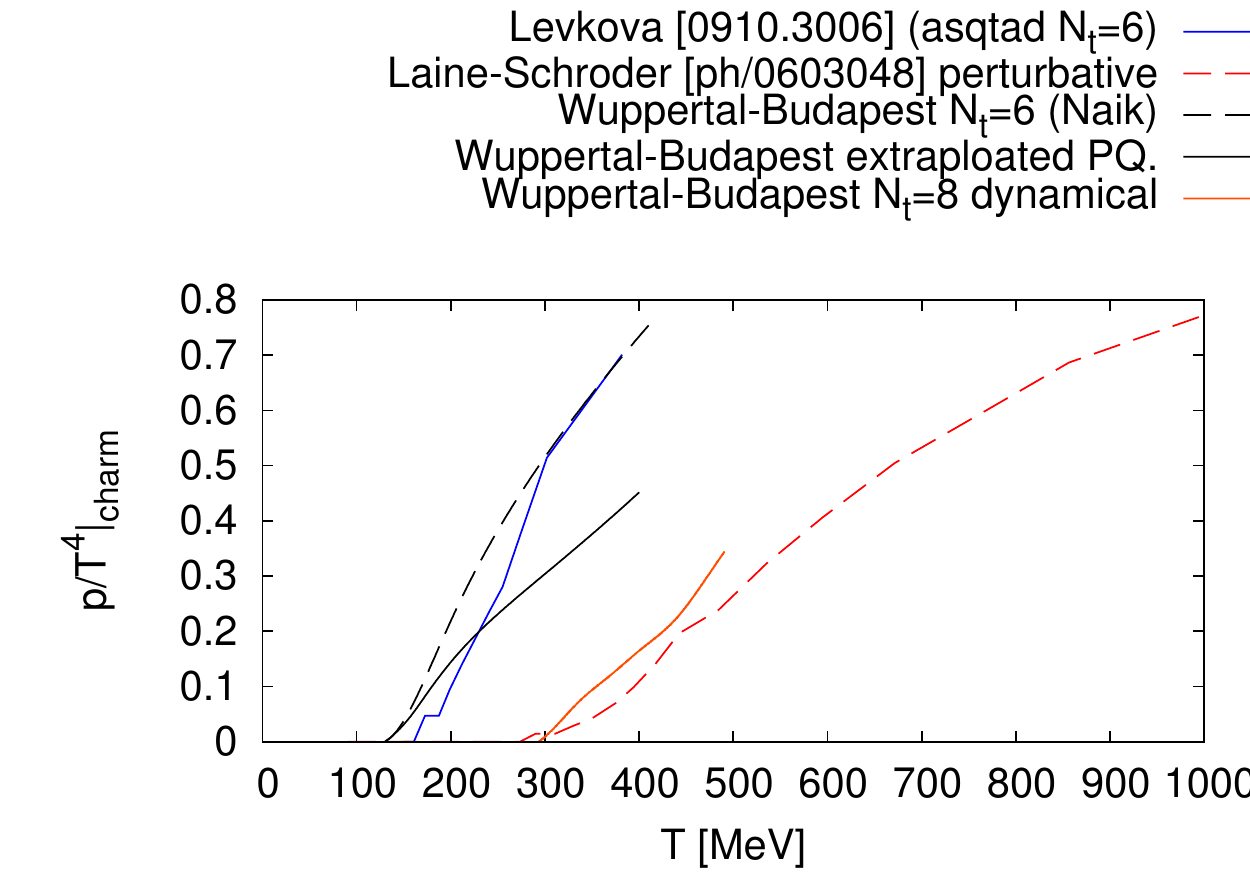}
\end{center}
\caption{\emph{Left panel:} Perturbative calculation~\cite{Laine:2006cp} of the pressure normalized by the temperature. In this perturbative estimate, the three and four flavor results start to deviate for temperatures around $300$~MeV. \emph{Right panel:} Summary of different calculations of the deviation of $N_f=2+1+1$ EOS from the $N_f=2+1$ EOS, both partially quenched and fully dynamical. From top down these results are: \cite{Levkova:2009gq}, \cite{Laine:2006cp} (corresponding to the result on the left panel), results from a partially quenched calculation including a ``Naik'' term, a continuum estimate for the partially quenched charm contribution, and a fully dynamical calculation at $N_t=8$, compared to the $N_f=2+1$ continuum estimate. Clearly, the partially quenched results consistently deviate already at low temperatures, whereas the dynamical results and the perturbative estimate both start to deviate at $T\sim 300$~MeV. }
\label{fig:pq}
\end{figure}
A straightforward approach in estimating charm quark effects is to include the charm quark in the valence sector only. Estimates of charm quark effects computed with this ansatz are discussed in \cite{Borsanyi:2010cj, Levkova:2009gq, Cheng:2007wu}. Here, the charm quark becomes relevant already for temperatures around 200~MeV, which is incompatible with perturbative estimates \cite{Laine:2006cp}, see Figure~\ref{fig:pq}.

\begin{figure}
\begin{center}
\includegraphics*[width=.6\textwidth]{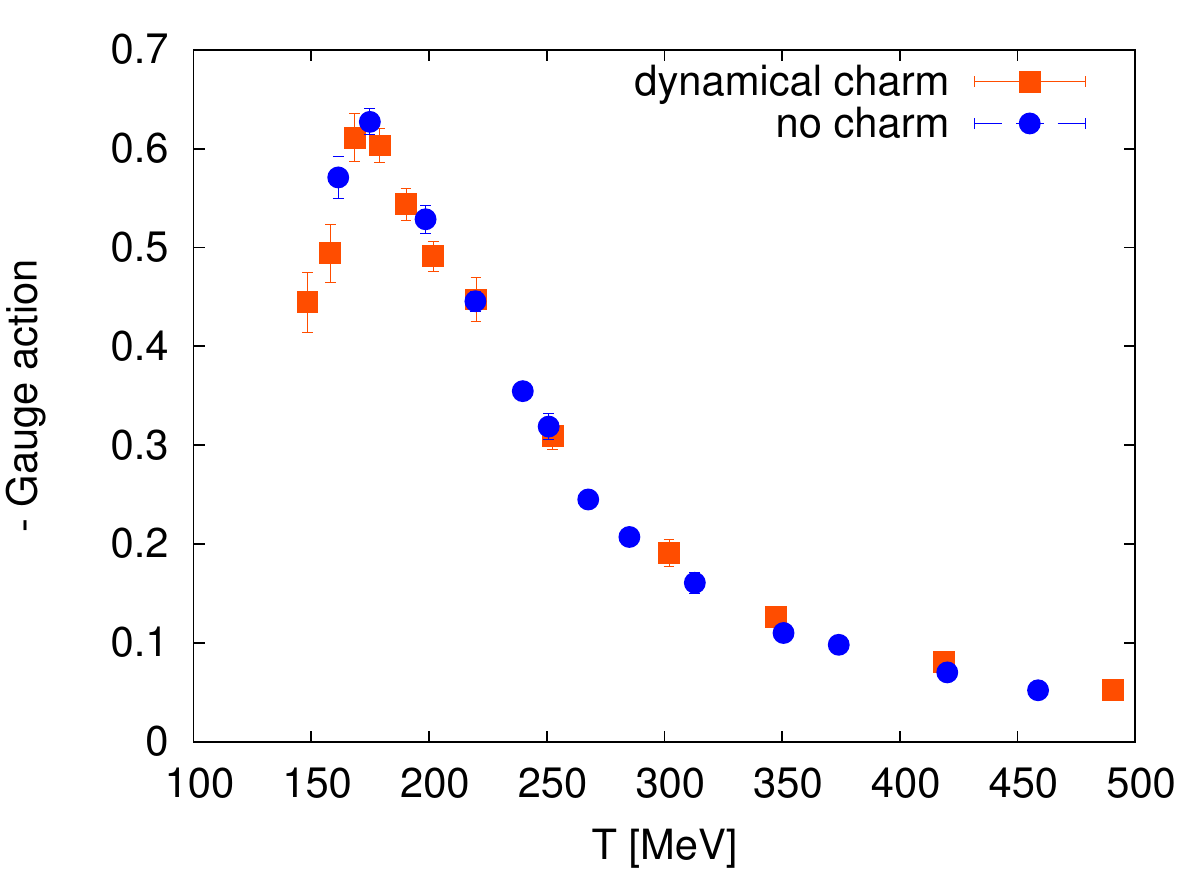}
\end{center}
\caption{Subtracted negative gauge action ($\langle-s_g\rangle^{sub}$ in the language of \cite{Borsanyi:2010cj}), used as input in the integral technique for calculating the EOS. Shown is the gauge action from the $N_f=2+1$ set, as it would apply to a partially quenched calculation, compared to the preliminary results for $N_f=2+1+1$. In terms of physical units, no sizable effect is visible.}
\label{fig:shift}
\end{figure}
\begin{figure}
\begin{center}
\includegraphics*[width=.48\textwidth]{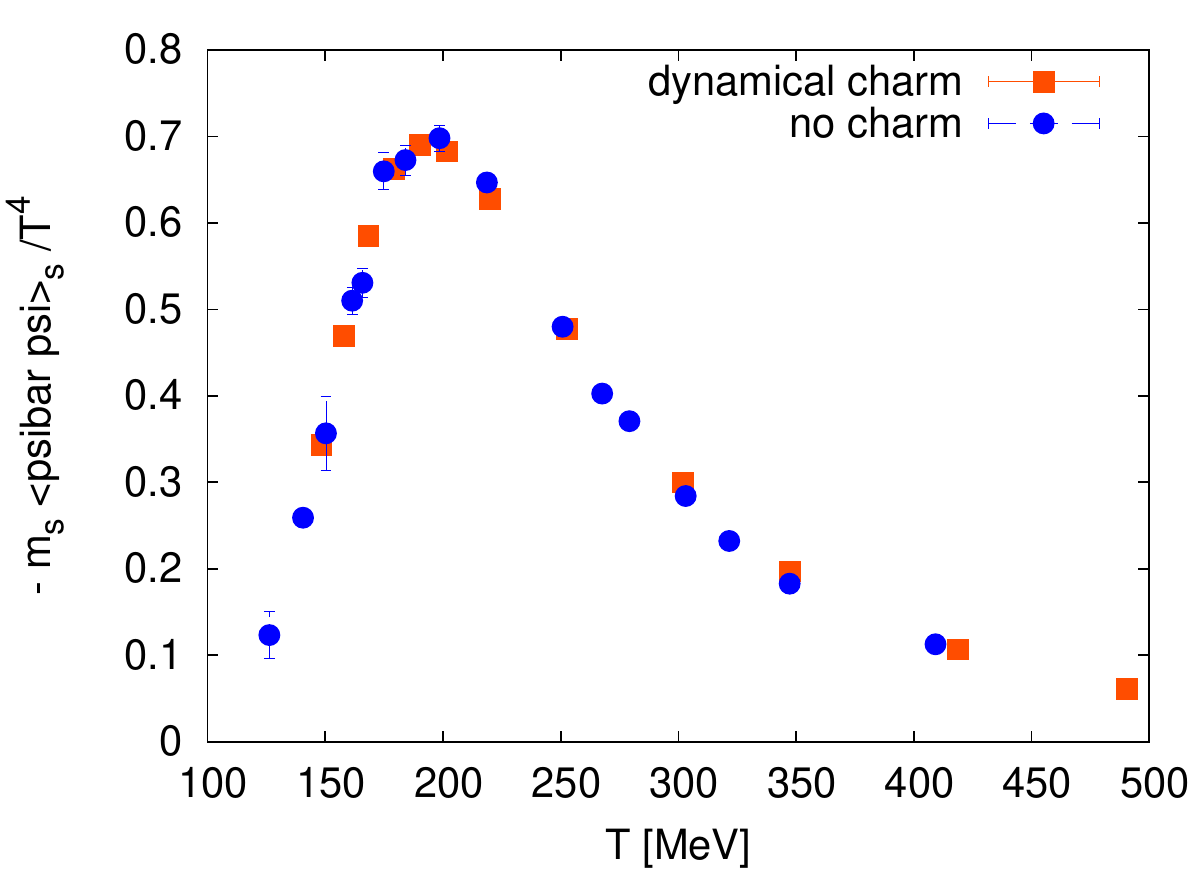}
\hspace{.02\textwidth}
\includegraphics*[width=.48\textwidth]{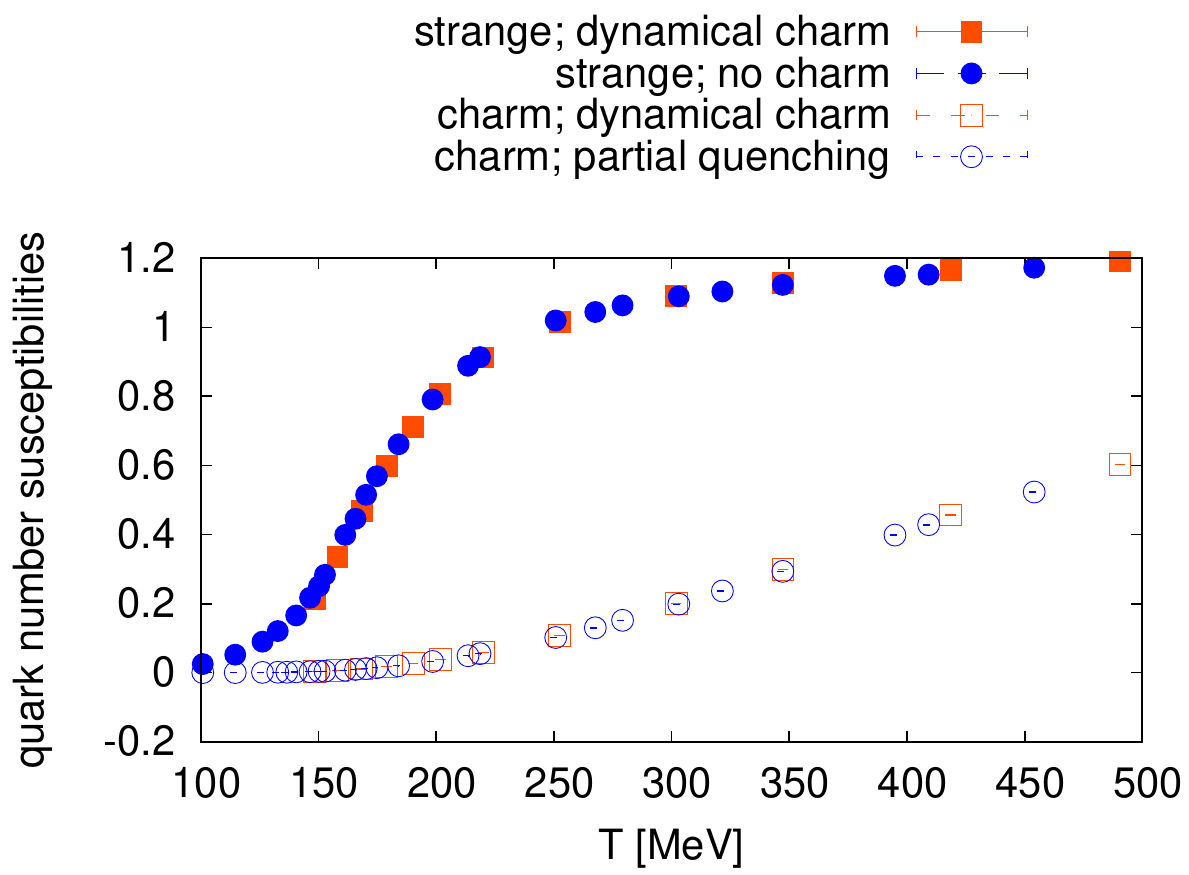}
\end{center}
\caption{In terms of physical units, the effects of the dynamical charm quark are small. \emph{Left panel:} Subtracted strange condensate. \emph{Right panel:} Quark number susceptibilities. }
\label{fig:shift2}
\end{figure}
This is analyzed more closely in Figure~\ref{fig:shift}. Here, the gauge action is plotted in terms of the temperature as defined through the LCP for both $N_f=2+1$ and $N_f=2+1+1$ sea quarks. 
Note that the gauge action enters into the calculation of the equation of state through the integral technique (here, $p$ refers to the pressure, and $\langle\rangle^{sub}$ to subtracted/renormalized quantities; our notation is explained in \cite{Borsanyi:2010cj}):
\begin{equation}
\label{eq:intp}
\frac{p(T)}{T^4}-\frac{p(T_0)}{T_0^4}=
N_t^4 \int^{(\beta,m_q)}_{(\beta_0,m_{q0})}
\left(d\beta \langle -s_g \rangle^{\rm sub} +
\sum_q d m_q \langle \bar\psi_q\psi_q\rangle^{\rm sub}\right).
\end{equation}
In terms of physical units, no sizable effects from the charm quark in the sea are visible, as is the case for other quantities as well (see Figure~\ref{fig:shift2}). 

The large deviation between the results shown in Figure~\ref{fig:pq} is, therefore, likely due to a shift in the LCP, which enters into eq.~\ref{eq:intp} through the measure. It is conceivable to try to correct for this shift using an $N_f=2+1+1$ LCP, but in the full result some effects will remain. Hence, we prefer to perform a fully dynamical simulation of the charmed EOS.

\section{Charmed equation of state for QCD}
\begin{figure}
\begin{center}
\includegraphics*[width=.75\textwidth]{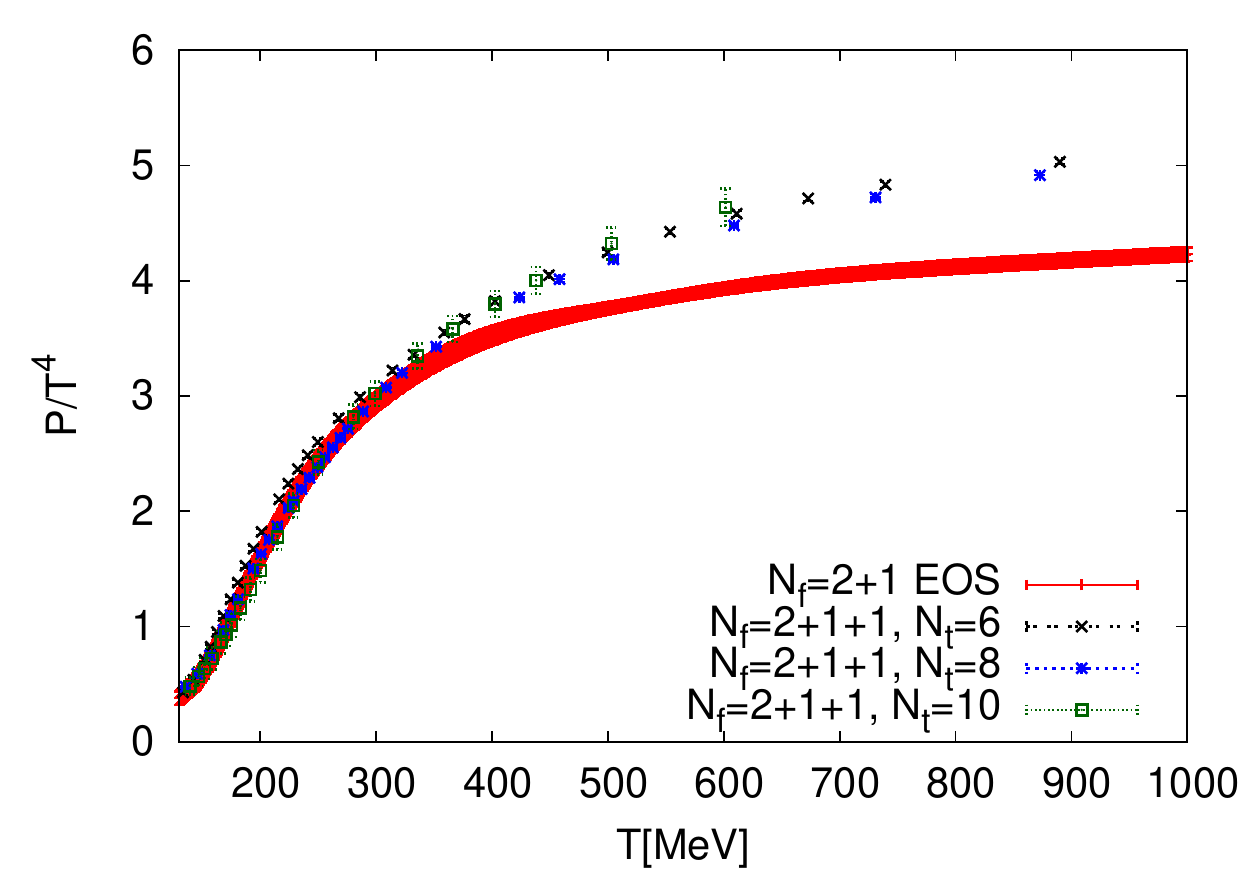}
\end{center}
\caption{$N_f=2+1+1$ EOS. The data point are tree-level corrected and compared to the $N_f=2+1$ continuum estimate. The $N_f=2+1+1$ and $N_f=2+1$ EOS agree well up to a temperature of about 300~MeV.}
\label{fig:eos}
\end{figure}
Using a new $N_f=2+1+1$ LCP, we computed the EOS for QCD for $N_t=6,8$, and $10$, as shown in Figure~\ref{fig:eos}. As already mentioned earlier, the fully dynamical charmed EOS agrees with the $N_f=2+1$ EOS up to higher temperatures than the partially quenched one. Similarly to the perturbative estimate, the temperature where charm effects become sizeable is close to $300$~MeV. 

Furthermore, there do not seem to be sizeable discretization artifacts due to the heavy charm. At low temperatures, where the lattice spacing is coarse, the data points for the different $N_t$ agree with each other, as well as with the $N_f=2+1$ EOS. As the temperature increases, and the lattice spacing becomes increasingly finer, potential discretization effects should become smaller. At our present level of precision, we do not see sizable deviations between the different $N_t$ from low to high temperatures; therefore, we presently believe that discretization effects due to the heavy charm quark will not be significant, or at least minor.

\section{Conclusion}
We have presented a first attempt to provide a continuum extrapolated equation of state for QCD with $N_f=2+1$ flavors of quarks, and have shown that it is free of finite size effects. We intend to supplement this calculation with additional $N_t=12$ and very few $N_t=16$ data points. 
In addition, we have addressed the issue of partially quenching for the (charmed) $N_f=2+1+1$ equation of state, and have argued that the difference of partially quenched and fully dynamical results is caused by a shift in the line of constant physics. Finally, we have shown first fully dynamical results for the charmed equation of state.

\section*{Acknowledgements}
Computations were performed on the Blue Gene supercomputers at Forschungszentrum J\"ulich, on clusters~\cite{Egri:2006zm} at Wuppertal and E\"otv\"os University, Budapest and on the QPACE at Forschungszentrum J\"ulich and Wuppertal University. This work is supported in part by the Deutsche Forschungsgemeinschaft grants FO 502/2 and SFB- TR 55 and by the EU (FP7/2007-2013)/ERC no. 208740.

\bibliography{refs}{}
\bibliographystyle{utphys}

\end{document}